\documentclass[fleqn,usenatbib]{mnras}

\usepackage{newtxtext,newtxmath}

\usepackage[T1]{fontenc}
\usepackage{ae,aecompl}

\usepackage{graphicx}	
\usepackage{amsmath}	
\usepackage{amssymb}	
\usepackage{color, colortbl}
\definecolor{Gray}{gray}{0.9}
\usepackage{longtable}
\usepackage{multicol}
\usepackage{multirow}

\title[Reflares of MAXI~J1535-571]{A \textit{NICER} look at the state transitions of the black hole candidate MAXI~J1535-571 during its reflares}

\author[V. A. C\'uneo et al.]{
V. A. C\'uneo,$^{1,2,3}$\thanks{E-mail: virginiacuneo@gmail.com}
K. Alabarta,$^{4,5}$
L. Zhang,$^{4}$
D. Altamirano,$^{4}$
M. M\'endez,$^{5}$
\newauthor 
M. Armas Padilla,$^{1,2}$
R. Remillard,$^{6}$
J. Homan,$^{7,8}$
J. F. Steiner,$^{9}$
J. A. Combi,$^{3,10}$
\newauthor 
T. Mu\~noz-Darias,$^{1,2}$
K. C. Gendreau,$^{11}$
Z. Arzoumanian,$^{11}$
A. L. Stevens,$^{12}$
\newauthor 
M. Loewenstein,$^{11,13}$
F. Tombesi$^{11, 13, 14}$,
P. Bult$^{11,13}$,
A. C. Fabian$^{15}$, 
\newauthor 
D. J. K. Buisson$^{4}$,
J. Neilsen$^{16}$,
and 
A. Basak$^{17}$
\\
$^{1}$Instituto de Astrof\'isica de Canarias (IAC), V\'ia L\'actea s/n, La Laguna 38205, S/C de Tenerife, Spain \\
$^{2}$Departamento de Astrof\'isica, Universidad de La Laguna, La Laguna, E-38205, S/C de Tenerife, Spain \\
$^{3}$Instituto Argentino de Radioastronom\'ia, (CCT - La Plata, CONICET - CICPBA), C.C.5, 1894 Villa Elisa, Buenos Aires, Argentina\\
$^{4}$Department of Physics and Astronomy, University of Southampton, Southampton, SO17 1BJ, UK\\
$^{5}$Kapteyn Astronomical Institute, University of Groningen, PO Box 800, NL-9700 AV Groningen, the Netherlands\\
$^{6}$MIT Kavli Institute for Astrophysics and Space Research, 70 Vassar Street, Cambridge, MA 02139, USA\\
$^{7}$Eureka Scientific, Inc., 2452 Delmer Street, Oakland, CA 94602, USA\\
$^{8}$SRON, Netherlands Institute for Space Research, Sorbonnelaan 2, 3584 CA Utrecht, The Netherlands\\
$^{9}$Center for Astrophysics, Harvard University, 60 Garden Street, Cambridge, MA 02138, USA\\
$^{10}$Facultad de Ciencias Astron\'omicas y Geof\'isicas, Universidad Nacional de La Plata, Paseo del Bosque, B1900FWA La Plata, Argentina\\
$^{11}$X-ray Astrophysics Laboratory, Astrophysics Science Division, NASA Goddard Space Flight Center, Greenbelt, MD 20771, USA\\
$^{12}$Department of Physics \& Astronomy, Michigan State University, 567 Wilson Road, East Lansing, MI48824, USA\\
$^{13}$Department of Astronomy, University of Maryland, College Park, MD 20742, USA\\
$^{14}$Department of Physics, University of Rome ``Tor Vergata", Via della Ricerca Scientifica 1, I-00133 Rome, Italy\\
$^{15}$Institute of Astronomy, Madingley Road, Cambridge, CB3 0HA, UK\\
$^{16}$Department of Physics, Villanova University, Villanova, PA 19085, USA\\
$^{17}$Anton Pannokoek Institute, University of Amsterdam, Science Park 904, 1098XH, Amsterdam, NL.\\
}

\date{Accepted XXX. Received YYY; in original form ZZZ}

\pubyear{2020}

\begin{document}
\label{firstpage}
\pagerange{\pageref{firstpage}$-$\pageref{lastpage}}
\maketitle

\begin{abstract}
The black hole candidate and X-ray binary MAXI~J1535-571 was discovered in September 2017. During the decay of its discovery outburst, and before returning to quiescence, the source underwent at least four reflaring events, with peak luminosities of $\sim$10$^{35-36}$ erg s$^{-1}$ (d/4.1\,kpc)$^2$. To investigate the nature of these flares, we analysed a sample of \textit{NICER} observations taken with almost daily cadence. In this work we present the detailed spectral and timing analysis of the evolution of the four reflares. The higher sensitivity of \textit{NICER} at lower energies, in comparison with other X-ray detectors, allowed us to constrain the disc component of the spectrum at $\sim$0.5 keV. We found that during each reflare the source appears to trace out a q-shaped track in the hardness-intensity diagram similar to those observed in black hole binaries during full outbursts. MAXI~J1535-571 transits between the hard state (valleys) and softer states (peaks) during these flares. Moreover, the Comptonised component is undetected at the peak of the first reflare, while the disc component is undetected during the valleys. Assuming the most likely distance of 4.1 kpc, we find that the hard-to-soft transitions take place at the lowest luminosities ever observed in a black hole transient, while the soft-to-hard transitions occur at some of the lowest luminosities ever reported for such systems. 
\end{abstract}




\section{Introduction}
Low mass X-ray binaries (LMXBs) are binary systems composed of a compact object, a neutron star or a black hole, that accretes material from the outer layers of a low-mass (usually $\leqslant$ 2 M$_{\odot}$) companion star. During this process, an accretion disc develops around the compact object and X-rays are emitted due to the high temperatures ($\sim$10$^{7}$ K) reached by the infalling material in the innermost parts. Most of these systems are transient, displaying outburst and quiescence periods. During outburst, the matter is accreted rapidly and the system reaches high X-ray luminosities ($L_X \sim$10$^{36-39}$ erg s$^{-1}$). Contrarily, in quiescence the accretion rate and X-ray luminosities are lower ($L_X \sim$10$^{31-34}$~erg~s$^{-1}$). Transient LMXBs spend most of their life time in the quiescent state, but they occasionally exhibit outbursts \citep[e.g.][]{2001NewAR..45..449L} that might last from some days \citep[see][]{2010ApJ...714..894H,2010ApJ...712L..58A} to several years \citep[for recent observational studies of X-ray binaries see, e.g,][]{2016A&A...587A..61C,2016ApJS..222...15T}. Outbursts are mainly characterised by two spectral states: hard and soft \citep[e.g.][]{2006ARA&A..44...49R,2013PASJ...65...26M}. In the hard state the X-ray spectrum is dominated by the emission of Comptonisation from a hot corona \citep[e.g.][]{2007A&ARv..15....1D,2017MNRAS.466..194B} and it can be modelled by a power law. In the soft state, the X-ray spectrum is dominated by thermal emission from an optically thick, geometrically thin accretion disc \citep{1973A&A....24..337S}, modelled by a disc black-body. LMXBs usually transit these states describing a ``q-shaped" track, known as hysteresis loop, moving counter-clockwise \citep[e.g.][]{2005A&A...440..207B,2010MNRAS.403...61D,2011BASI...39..409B,2016ASSL..440...61B} in the hardness-intensity diagram \citep[HID;][]{2001ApJS..132..377H}. Although the HID is useful to trace the general evolution of the source, an analysis of the variability adds extra information, establishing e.g. the boundaries of the different states. In black hole systems the hard state is characterised by having a broadband fractional rms of $\sim$30$-$40 per cent, the hard intermediate $\sim$10$-$30 per cent, the soft intermediate $\sim$5$-$10 per cent, and the soft state $<$ 5 per cent \citep[][see also \citealp{1997ApJ...479..926M,2012MNRAS.427..595M,2012MNRAS.422.2620H}]{2011MNRAS.410..679M}.

Outbursts in transient LMXBs are generally characterised by a fast brightness rise, followed by a slower decay to quiescence. However, during the decaying phase, or when the source is reaching the quiescence state, sometimes the accretion rate rises again and the source rebrightens, reaching X-ray luminosities one or two orders of magnitude fainter than the outburst peak. These episodes, known in the literature as reflares, rebrightenings, rebursts, mini-outbursts or echo-outbursts, have been observed in outbursts of a limited number of sources \citep[e.g.][]{1995ApJ...441..786C,2011ApJ...742L..17A,2012MNRAS.423.3308J,2016ApJ...817..100P}. It is generally believed that the rebrightenings are ``echoes" of the main outburst, in the sense that they are also due to an increase in the mass accretion rate, resembling the main outburst physical properties and phenomena, but in a smaller scale. Moreover, given that they are very common in dwarf novae systems, it is believed they are related to the accretion process and the companion star more than to the type of compact object. However, it is still unknown what produces reflaring and why it manifests only for some sources and for some outbursts. The main issue for studying these events is that observing campaigns usually stop when the sources head to quiescence, and in many cases the rebrightenings are not observed. In addition, sometimes the rebrightenings are only observed at some spectral ranges of the X-ray or optical bands. This makes some instruments more sensitive to reflares, and their detection depends on which instrument is observing. Moreover, since rebrightenings are several orders of magnitude fainter than the main outburst, it is difficult to infer strong conclusions based on colour/hardness studies as the data are dominated by low statistics. Although what causes the increase of the mass accretion rate that give rise to a rebrightening is still unknown, improved monitoring from instruments such as \textit{Swift} \citep[Neil Gehrels Swift Observatory;][]{2004ApJ...611.1005G}, \textit{MAXI}, \textit{NICER} (Neutron star Interior Composition Explorer)\footnote{\url{https://heasarc.gsfc.nasa.gov/docs/nicer/}} and \textit{INTEGRAL}, have increased significantly our ability to detect them.

The X-ray binary transient MAXI~J1535-571 (J1535) was discovered simultaneously by \textit{MAXI} \citep{2017ATel10699....1N} and \textit{Swift} \citep{2017GCN.21788....1M,2017ATel10700....1K}, on September 2nd, 2017 (MJD 57999). The estimated peak luminosity in the 1$-$60~keV band and the observed rapid X-ray variability in \textit{MAXI} light curves (2$-$20~keV), suggested that J1535 is a LMXB harbouring a black hole \citep{2017ATel10708....1N}. Additionally, \citet{2017ATel10711....1R} used Australia Telescope Compact Array (\textit{ATCA}) data to estimate a 5 GHz radio luminosity of 2.053$\pm$0.008 $\times$ 10$^{30}$~erg~s$^{-1}$ (d/6.5 kpc)$^2$, consistent with a black hole accretor \citep[see also][]{2019ApJ...883..198R,2019MNRAS.487.4221S}. J1535 exhibited the expected state transitions, hard-to-soft and soft-to-hard, for a LMXB during outburst \citep{2017ATel10699....1N,2017ATel10708....1N,2017ATel11020....1S,2018ATel11611....1R}. The first rebrightening of J1535 was detected in mid May, 2018, by \citet{2018ATel11652....1P} at 0.5$-$10~keV, and the fits to these data suggested that the source may have transitioned to the soft state. A total detection of 5 reflaring episodes, during the outburst declining, was reported by \citet{2019ApJ...878L..28P} from \textit{Swift} and \textit{ATCA} data. The source was no longer detected since May 2019 \citep{2019ATel12780....1P}.

\textit{NICER} monitored J1535, with almost daily cadence, from its discovery in 2017 until May, 2019, with the exception of the 2 periods when the source was behind the Sun [from 13 October 2017 to 5 January 2018 (MJD 58039$-$58123) and from 8 October 2018 to 10 January 2019 (MJD 58399$-$58493)]. In this paper we analyse \textit{NICER} data during the decaying phase. We focus on the sequence of rebrightenings observed from mid May until October, 2018. According to the classification of \citet{2019ApJ...876....5Z} for different types of rebrightenings, the events in J1535 are reflares, which is the term we will use from now on. However, \citet{2019ApJ...878L..28P} consider that the source had returned to quiescence before the rebrightening episodes, making J1535 to lie outside of this classification.

In Section 2 we describe the observations and data analysis. Results from the analysis of the light curve, hardness ratio, energy spectra and power spectra, are reported in Section 3. In Section 4 we discuss the implications of our results and compare J1535 reflares with similar events exhibited by other sources. We also discuss the possible scenarios for the origin of reflaring events. Conclusions follow in Section 5.

\section{Observations and data analysis}
\subsection{\textit{NICER}}
\subsubsection{Data}
\textit{NICER} on-board the International Space Station is an observatory that observes in the 0.2$-$12 keV band, with a time resolution $<$ 300 nsec \citep{2012SPIE.8443E..13G}. \textit{NICER}'s X-ray Timing Instrument monitored J1535 since its discovery in September, 2017. 
As we are interested in the reflares observed at the end of the outburst, in this paper we focus our analysis on observations taken within MJD 58250$-$58398 (ObsID: 1130360177$-$1130360276). We removed data from detectors \#14 and \#34, which occasionally show episodes of increased electronic noise. The data were processed using \textsc{nicerdas} version 6.0 and the \textsc{caldb} version 20190516. We applied standard filtering and cleaning criteria to obtain the cleaned event files, including data with a pointing offset of $<54\arcsec$, bright Earth limb angle $>40\degr$, dark Earth limb angle $>30\degr$, and outside the South Atlantic Anomaly. Furthermore, we excluded time intervals showing strong background flare-ups. The background was calculated using the ``3C50\_RGv5" model provided by the \textit{NICER} team. Individual \textit{NICER} observations have a typical duration of a few ksec that were subdivided into several data segments, separated by 1$-$3 ksec gaps due to the orbit of the ISS and composed of good time intervals.

\subsubsection{Energy spectra}
We extracted a background-subtracted energy spectrum for each data segment during the reflaring periods. We rebinned the spectra by a factor of 3 to correct for energy oversampling and then grouped them in bins of at least 30 counts. The obtained energy spectra were then fitted in the 0.5$-$10~keV band using \textsc{xspec} version 12.10.1, within \textsc{heasarc}. We applied version 1.02 of the ancillary response file and version 1 of the response matrix, both provided by the \textit{NICER} team. A preliminary analysis of the spectra showed the presence of a disc component and a Comptonisation component, as it is usual for black hole LMXBs. We fitted every spectrum with an absorbed 2-component model: a thermally Comptonised continuum component \citep{1996MNRAS.283..193Z,1999MNRAS.309..561Z} plus a multi-temperature disc component [\textsc{tbabs*(nthcomp+diskbb)}]. The choice of \textsc{nthcomp} over other Comptonisation models relies on the better description of a thermal Comptonisation due to the ability to constrain the low and high energies rollover. The interstellar absorption was modelled using the Tuebingen-Boulder intestellar absorption model (\textsc{tbabs}), with solar abundances set according to \citet{2000ApJ...542..914W} and the cross-sections from \citet{1996ApJ...465..487V}. The hydrogen column density was fixed to 3.2 $\times$ 10$^{22}$ cm$^{-2}$, the resulting value from fitting the spectrum in our sample with the highest net counts number. This value is consistent with previous estimates \citep{2018MNRAS.480.4443T,2017ATel10700....1K,2018ApJ...868...71S,2019ApJ...875....4S}. Given that the cutoff energy of the Comptonised component lies outside of our energy range, we fixed the electron temperature. We chose 100 keV due to the likely black hole nature of the compact object \citep[see][]{2017MNRAS.466..194B}. The seed photon temperature was linked to the disc temperature. We estimated 1$\sigma$ errors for all the spectral parameters. X-ray luminosities were estimated from the unabsorbed flux, calculated with the \textsc{cflux} convolution model within \textsc{xspec} (0.5$-$10 keV), assuming a distance of 4.1$^{+0.6}_{-0.5}$ kpc \citep{2019MNRAS.488L.129C}, a value we adopt from now on.   

During a period of 3 days (MJDs 58259, 58260 and 58261;  corresponding to the peak of the first reflare), the background model underestimated the background count rate at $E>6$ keV for some of the orbits. The  energy spectra of the source was dominated by the low-energy component during these observations, and the photon index $\Gamma$ of the Comptonised component was unconstrained due to the background uncertainties. Therefore we only report the best fit parameters for the disc component during these 3 days. 

Mostly during the valleys of the reflares, we observed that the normalisation of the disc component was not 3$\sigma$ significant. In order to determine whether a disc component (detected at the peaks) is present also in the valleys, we estimated upper limits. In all cases, we fixed the disc temperature to 0.49 keV, the value obtained in the last fit where the component was significant from the first reflare (where the evidence of the presence of the disc is stronger). An analogous situation occurred for some fits after the peak of reflare 1, but for the Comptonised component. In these cases, we fixed $\Gamma$ at 2. We also considered as upper limits all the parameters that were consistent with 0 within errors.

\subsubsection{Light curve and hardness-intensity diagram}
\label{lc-hid}
We produced a light curve from the count rate in the 0.5$-$10~keV band, using \textsc{xspec}. We defined hardness as the ratio between the 3$-$10 keV and the 0.5$-$3 keV count rates.

\begin{figure*}
 \includegraphics[trim=27mm 21mm 35mm 35mm,clip,width=\textwidth]{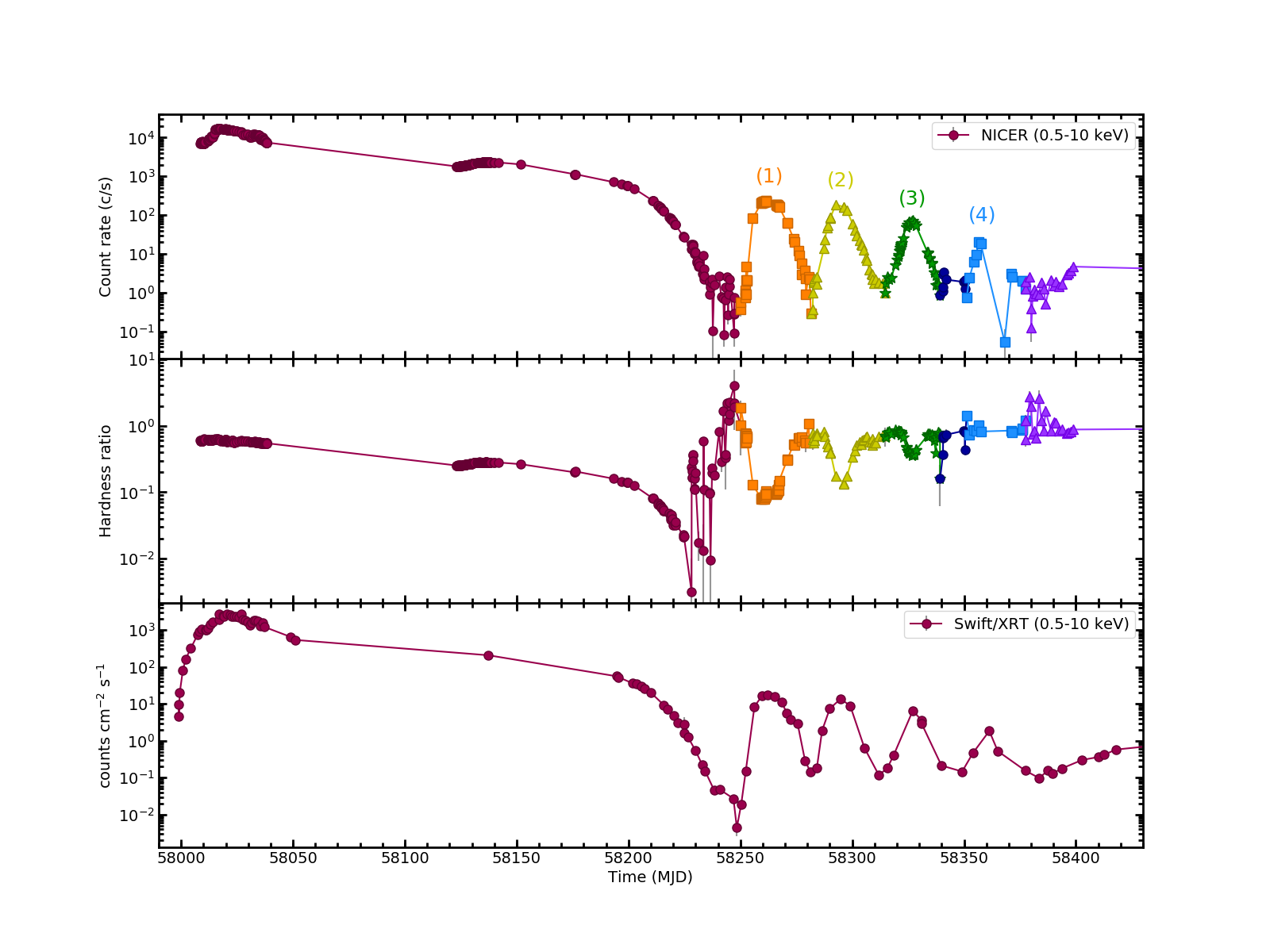}
 \caption{Long-term light curve of MAXI~J1535-571. Top panel: \textit{NICER} light curve in the 0.5$-$10 keV band. Each data point corresponds to the average count rate of a 1-orbit data segment. The colours denote the different reflares. The dark-blue portion corresponds to a undefined period between reflares 3 and 4 (see text for the definition of the onset of a reflare). The purple portion denotes a period of variation around low count rates and the possible onset of a fifth reflare to the end. Middle panel: Hardness evolution from \textit{NICER} data, defined as the ratio between the hard band (3$-$10 keV) and the soft band (0.5$-$3 keV). Bottom panel: Daily averaged \textit{Swift/XRT} light curve in the 0.5$-$10 keV band.}
 \label{fig:maxij1535}
\end{figure*}

\subsubsection{Timing analysis}
We calculated the Fast Fourier Transform (FFT) of the data. We generated the power spectra for each observation in the 0.5$-$10 keV band, using a minimum frequency of 7.6~$\times$~10$^{-3}$~Hz, a Nyquist frequency of 4000 Hz and data segments of $\sim$131 s. After visual inspection of each power spectra light curve, we cleaned the resulting power spectra by removing the data segments evidently produced by background features. We subtracted the Poisson noise from each power spectra. Finally, we renormalised by rms, using an average background count-rate for each observation estimated from the background energy spectra. We estimated the integrated fractional rms amplitude between the minimum frequency, 7.6 $\times$ 10$^{-3}$ Hz, and 10 Hz for those observations with a net count rate higher than 10 cts s$^{-1}$ in the 0.5$-$10~keV band. Some of the estimated rms amplitudes were not 3$\sigma$ significant individually. We therefore combined the power spectra of groups of non-significant rms amplitude that were close in time and had a similar hardness. Some of the averages were still not significant, but we nevertheless included them in the analysis for consistency, with their 1$\sigma$ error (see Section \ref{timing}). In addition, we averaged all the power spectra of those observations with count rates between 1 and 10 cts s$^{-1}$. Finally, we computed the absolute rms (RMS) by multiplying the fractional rms by the net count rate in the 0.5$-$10 keV band.

\subsection{Swift}
For a comparison, we used data from the X-ray Telescope \citep[XRT;][]{2003SPIE.4851.1320B} on board \textit{Swift}. We used the light-curve generator\footnote{\url{http://www.swift.ac.uk/user_objects/}} provided by the UK Swift Science Data Centre \citep[][]{2007A&A...469..379E} to extract the long-term day averaged \textit{Swift/XRT} light curve of J1535, in the 0.5$-$10~keV band.

\section{Results}

\subsection{Light curve and hardness-intensity diagram}
Figure \ref{fig:maxij1535}, shows the long-term light curve of J1535. 
The top and bottom panels show the \textit{NICER} and the \textit{Swift/XRT} light curves, respectively. \textit{NICER} started the observations at MJD 58003, when the outburst of J1535 was still rising. The source reached its peak brightness around MJD 58017, with a luminosity of $\sim$4 $\times$ 10$^{38}$ erg s$^{-1}$. Despite the observational gap between MJD 58039 and 58123 due to the source field of view going behind the Sun, the brightness appeared to have declined slowly for $\sim$106 days, before showing a slight increase from MJD 58123. The source reached its second maximum at MJD 58136, with a luminosity of $\sim$5~$\times$~10$^{37}$~erg~s$^{-1}$, and then continued declining. On MJD 58202 the flux started to decrease more rapidly, until J1535 reached a minimum on MJD $\sim$58236 and fluctuated slightly at low luminosities ($\sim$10$^{34}$ erg s$^{-1}$) for $\sim$14 days. Starting from MJD $\sim$58250 the source exhibited a sequence of 4 reflares indicated from 1 to 4 in orange, yellow, green and light blue colours in time order, respectively, in Figure \ref{fig:maxij1535}. 
We consider that each reflare started when the slope of the light curve changed and began to increase, keeping the trend, after a decreasing period.
These flares occurred with an approximately averaged periodicity of $\sim$31$-$32 days.
Table \ref{reflares} summarises the main characteristics of the reflares. Due to a sampling difference between \textit{NICER} and \textit{Swift/XRT} observations, the dates are approximate. The peak luminosities ranged between 0.8 and 7 $\times$ 10$^{36}$~erg~s$^{-1}$, decreasing with time. After reflare 4, the count rate remained at a low value, similar to that reached before the reflaring events, at an average luminosity of $\sim$10$^{34}$ erg s$^{-1}$ for $\sim$15 days. In the last $\sim$3 days before the 2018 Sun constraint, the count rate started to increase again, strongly suggesting the onset of a fifth reflare \citep[see][]{2019ApJ...878L..28P}. The overall behaviour of the \textit{Swift/XRT} light curve of J1535 (bottom panel of Figure \ref{fig:maxij1535}) is, as expected, very similar to the \textit{NICER} light curve, although with a poorer sampling.

\begin{table*}
\caption{Summary of the main characteristics of the reflares}
\begin{tabular}{cccccc}
\hline 
Reflare & Start date & Duration & Peak date & Unabsorbed Peak & Peak luminosity \\
 & [MJD] & [Days] & [MJD] & Flux [erg cm$^{-2}$ s$^{-1}$] & [erg s$^{-1}$] \\
\hline
\rule{0pt}{3ex}1 & $\sim$58250 & $\sim$31 & $\sim$58261 & 3.37$\pm$0.02 $\times$ 10$^{-9}$ & 7$\pm$1 $\times$ 10$^{36}$ \\
\rule{0pt}{3ex}2 & $\sim$58282 & $\sim$32 & $\sim$58292 & 2.93$\pm$0.02 $\times$ 10$^{-9}$ & 6$\pm$1 $\times$ 10$^{36}$ \\
\rule{0pt}{3ex}3 & $\sim$58314 & $\sim$25$-$37 & $\sim$58327 & 1.33$\pm$0.02 $\times$ 10$^{-9}$ & 2.7$^{+0.6}_{-0.5}$ $\times$ 10$^{36}$ \\
\rule{0pt}{3ex}4 & $\sim$58340$-$52 & $\sim$31$-$43 & $\sim$58356 & 4.0$^{+2.3}_{-0.5}$ $\times$ 10$^{-10}$ & 8$^{+5}_{-2}$ $\times$ 10$^{35}$ \\
\noalign{\vskip 1mm}
\hline
\end{tabular}
\label{reflares}
\end{table*}

\begin{figure}
 \includegraphics[trim=3mm 0mm 15mm 10mm,clip,width=\columnwidth]{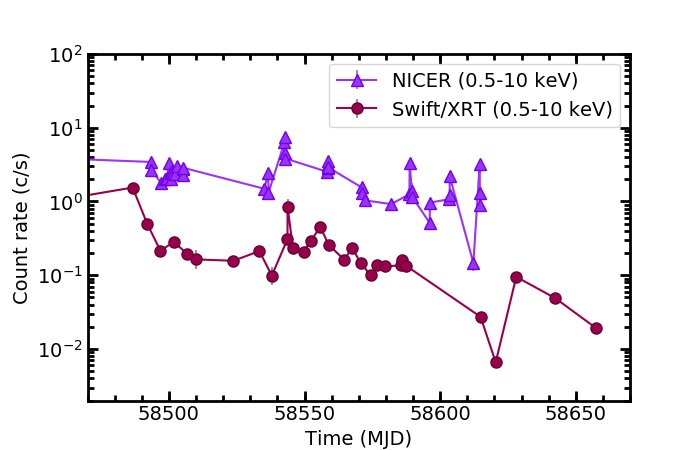} 
 \caption{\textit{NICER} (0.5$-$10 keV) and \textit{Swift/XRT} (0.5$-$10 keV) light curves from January 2019 until the end of both missions monitoring in May 2019. Error intensities are inside the symbols. The slow decline of the curves suggests the return to quiescence.}
 \label{after_sun}
\end{figure}

Figure \ref{after_sun} shows the light curve of the last part of the outburst, from MJD $\sim$58485 (January 2019), when J1535 was visible again after the Sun constraint, until the end of the monitoring (May 2019).

\begin{figure}
 \includegraphics[trim=4.5mm 30mm 12mm 37mm,clip,width=\columnwidth]{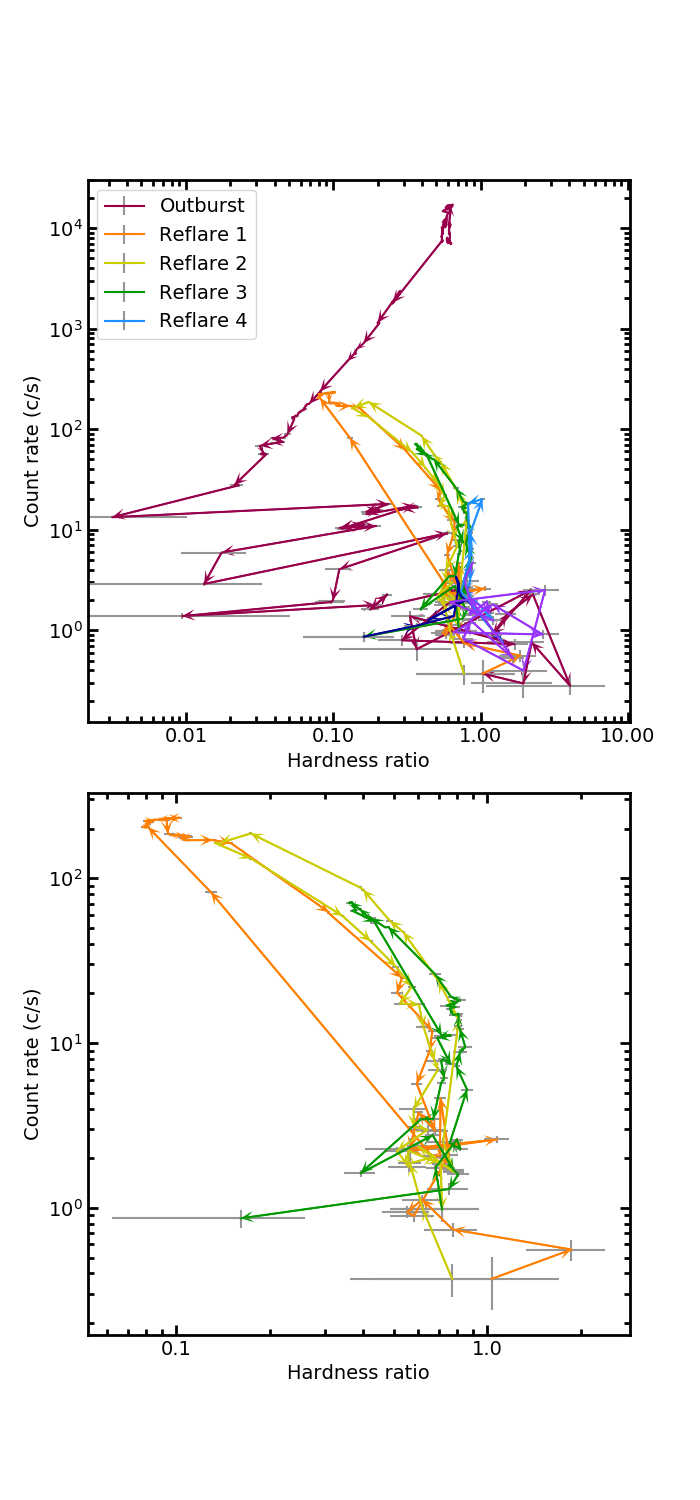}
 \caption{Top panel: Hardness-intensity diagram for MAXI~J1535-571. The arrows represent the direction of the evolution. The colours correspond to the outburst and reflares as in Figure \ref{fig:maxij1535}. Bottom panel: the same than in the top panel but for the bright reflares: 1, 2 and 3.}
 \label{fig:hid}
\end{figure}

The middle panel of Figure \ref{fig:maxij1535} shows the evolution of the hardness ratio. Before MJD $\sim$58228, the hardness decreases nearly monotonically to low values, following mainly the main outburst decline stage. The behaviour changes from MJD $\sim$58228. The source's intensity falls to low values and errors become large, while the hardness displays an increasing tendency, but also with large errors. On MJD $\sim$58247, right before the reflaring episodes start, the hardness is at a peak value of $\sim$4. The subsequent evolution is exactly opposite to the light curve evolution: the hardness is at a minimum when the source intensity peaks, and vice-versa. After reflare 4, when the intensity remains at low values, the hardness fluctuates around a value of $\sim$0.85. To study the state of the source with more detail, we plotted in the top panel of Figure \ref{fig:hid} the HID for J1535. Although the rising part of the outburst is missing, we observe that for the spectral bands we chose to define the hardness, J1535 does not trace the expected ``q-shape" during the outburst exactly, but it still evolves counter-clockwise, from the bright hard state to the soft state and back to the hard state. The source peaks at low values of the hard colour and then moves to soft colours while decreasing its intensity. 
After the source reaches its softest point at MJD $\sim$58228, we observe fluctuations in the hardness, however we note that the errors in those data points are large.
The source evolves then to harder values. During the reflares, J1535 also evolves counter-clockwise with the exception of reflare 1, in which it appears to evolve in the opposite direction. However, we note that this conclusion is just motivated  by a single data point and that the soft-to-hard transition takes place in a similar track than for reflares 2 and 3. 
The bottom panel of Figure \ref{fig:hid} shows the HID for the reflares. We observe that instead of completing the ``q-shape", the reflares fluctuate between the hard area and a softer region of the diagram, exhibiting hysteresis. However, during reflare 4 (in light blue at the left panel of Figure \ref{fig:hid}) the count rates are too low to detect such behaviour. The observed fluctuations of the source during the reflares point to smaller scale state transitions between a hard state and a ``softer" state.

\subsection{Energy spectra}
\begin{figure}
 \includegraphics[trim=2mm 6mm 14mm 13mm,clip,width=\columnwidth]{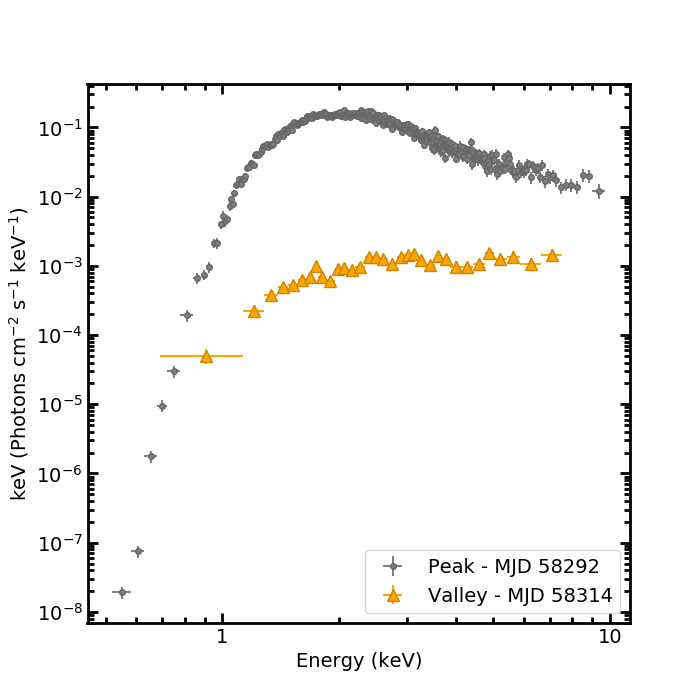}
 \caption{Representative energy spectra during the reflares, unfolded using the best fitting model in each case. The grey dots correspond to an observation near the peak of reflare 2, where both disc and Comptonisation models were used, while the orange triangles represent an observation in the valley between reflares 2 and 3, where the Comptonisation model was enough to fit the data.}
 \label{fig:spec_ex}
\end{figure}

\begin{figure*}
 \includegraphics[trim=21mm 38mm 33mm 40mm,clip,width=\textwidth]{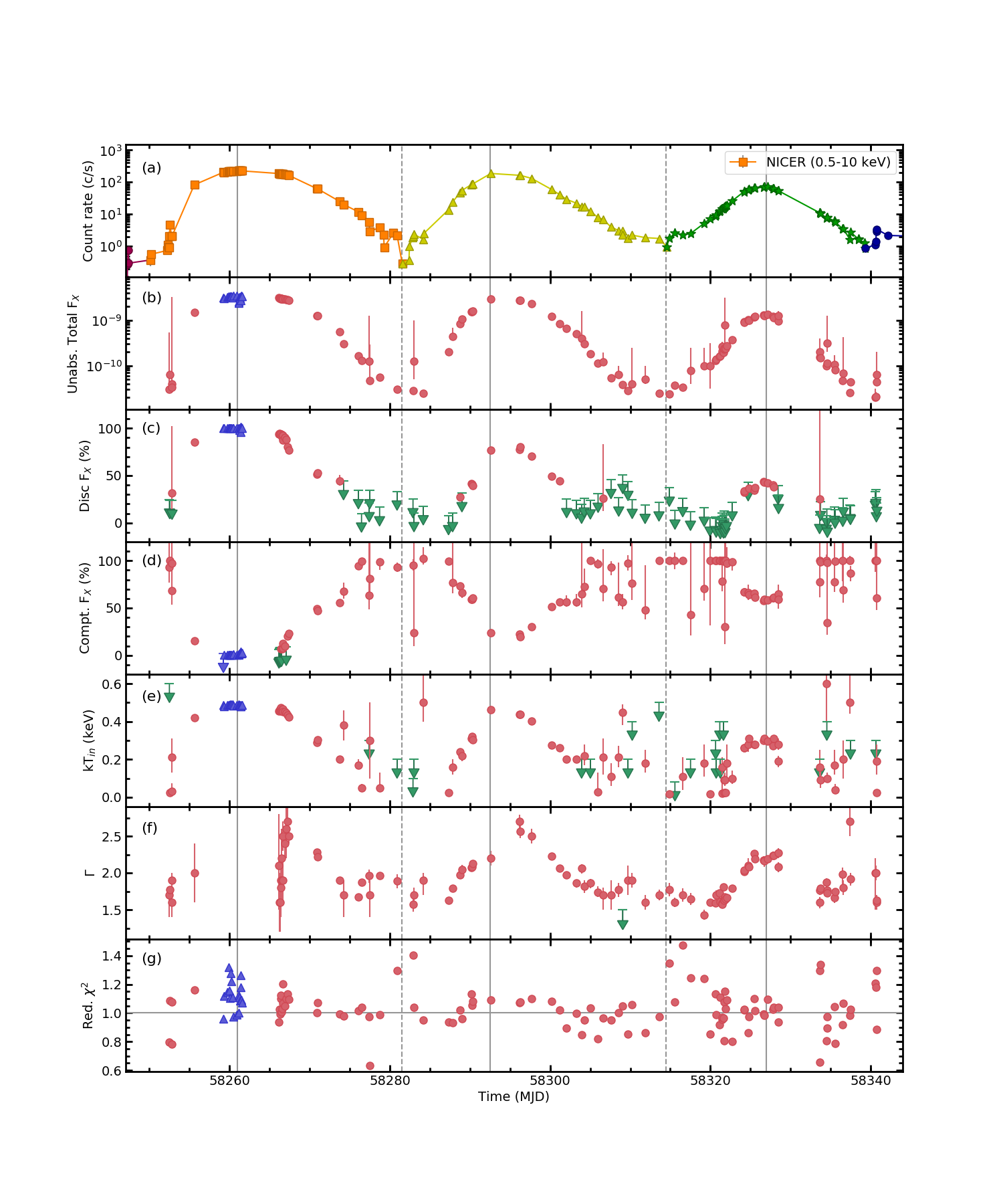}
 \caption{Evolution of the parameters obtained from the energy spectral fits using the model \textsc{tbabs*(nthcomp+diskbb)}, during the reflares. For a comparison, the top panel shows the light curve. The unabsorbed flux is in units of erg cm$^{-2}$ s$^{-1}$. The blue triangles correspond to the fits where we could not constrain $\Gamma$ (see Section 2.1.2 for more details). The green arrows correspond to the 90 per cent upper limits of spectra where the disc or the Comptonisation, respectively, were not 3$\sigma$ significant, or to those parameters consistent with 0 within errors. Grey vertical lines denote the peaks (solid) and valleys (dashed) of reflares.}
 \label{fig:spec_param}
\end{figure*}

Figure \ref{fig:spec_ex} shows representative energy spectra: one near the peak of reflare 2 and one in the valley between reflares 2 and 3. Figure \ref{fig:spec_param} shows the evolution of the parameters from the spectral fits. We only show the results for the first 3 reflares, as the fits for reflare 4 are not statistically sufficient. The panel (a) of Figure \ref{fig:spec_param} shows the \textit{NICER} light curve in units of count rates, while panel (b) is in units of unabsorbed flux (0.5$-$10 keV). Panels (c) and (d) show the percentage of the total flux corresponding to the disc and Comptonised components, respectively. While at the peaks the dominating component is the \textsc{diskbb}, the \textsc{nthcomp} component dominates the spectra when the source is in the valleys of the reflares. The temperature of the disc blackbody kT$_{in}$ is displayed in panel (e). We find that the temperature is correlated with the light curve, reaching values of $\sim$0.55 keV. Panel (f) in Figure \ref{fig:spec_param} shows the photon index $\Gamma$ evolution. We observe that $\Gamma$ also follows the light curve sinusoidal behaviour, being on average $\sim$2.7 at the peak of the reflares and $\sim$1.5 in the valleys. Finally, panel (g) of Figure \ref{fig:spec_param} shows the reduced $\chi^2$. The best-fitting spectral parameters are compiled and available in a Table as online-only material. 

\subsection{X-ray variability}
\label{timing}
\begin{figure}
 \includegraphics[trim=5mm 26mm 13mm 36mm,clip,width=\columnwidth]{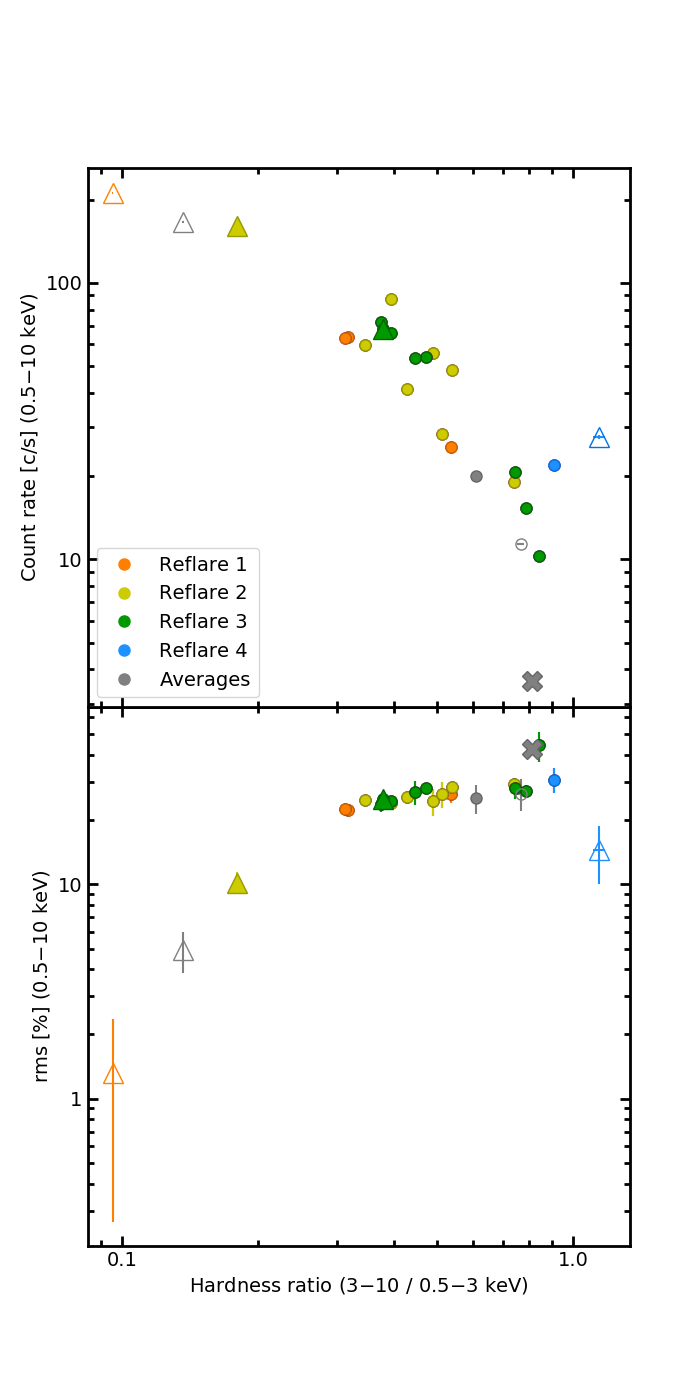}
 \caption{Top panel: HID during the reflaring epoch for the observations with a count rate higher than 10 cts s$^{-1}$. The cross is the average of all low count rate observations. The triangles denote averages of observations at the peak of the 4 reflares, plus a point that is an average of observations at the peak of reflares 1 and 2. The dots are individual or averaged observations not in the peaks. The grey symbols are averages of observations during more than one reflare. Bottom panel: broadband rms amplitude in the 7.6~$\times$~10$^{-3}$$-$~10~Hz frequency range as function of the hardness ratio for the same data set as in the top panel. The open symbols in both panels correspond to individual or averaged power spectra with rms amplitudes that were less than 3$\sigma$ significant.}
 \label{fig:rvh}
\end{figure}

Figure \ref{fig:rvh} shows the HID (top panel) and the hardness-rms (7.6 $\times$ 10$^{-3}$ $-$ 10 Hz) diagram \citep[bottom panel; e.g.][]{2005A&A...440..207B} from MJD 58250 onward. The bottom panel of Figure \ref{fig:rvh} shows that the rms evolves from $\sim$1.3 per cent, at a hardness value of $\sim$0.09, to $\sim$30 per cent, at a hardness of $\sim$0.9, with the exception of three data points. 

\begin{figure}
 \includegraphics[trim=5mm 5mm 13mm 15mm,clip,width=\columnwidth]{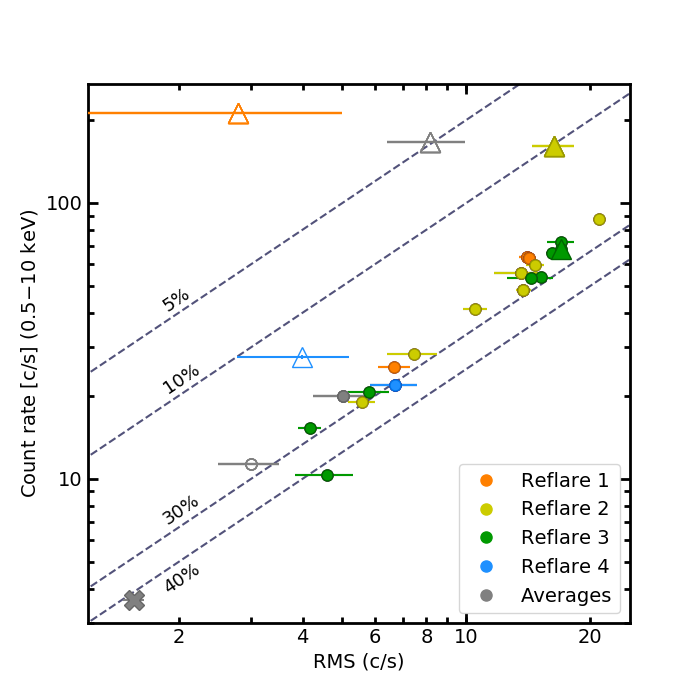}
 \caption{RMS vs count rate in the 0.5$-$10 keV band for the same data set as in Figure \ref{fig:rvh}. The dashed lines denote the 5 per cent, 10 per cent, 30 per cent and 40 per cent lines of constant fractional rms.}
 \label{fig:rvi}
\end{figure}

Figure \ref{fig:rvi} shows the RMS vs. the net count rate. We observe that most of the data are located in, or slightly to the right, of the $\sim$30 per cent of the rms, which was defined in the literature as the ``hard line" \citep[][but note that these authors used wider energy and frequency bands to measure the rms amplitude]{2011MNRAS.410..679M}. 
The three data points at a count rate higher than 100 cts s$^{-1}$ correspond to the observations at the peaks of reflares 1 and 2: the point at an rms $<$ 5 per cent is an average of observations at the peak of reflare 1 (orange triangle), the point at $\sim$10 per cent is an average of observations at the peak of reflare 2 (yellow triangle), and the point at $\sim$5 per cent is an average of observations at both peaks (grey triangle).

The observations lying along the ``hard line", at count rates lower than $\sim$100 cts s$^{-1}$, are associated with energy spectra with a dominant Comptonised component, evidencing that the source was effectively in the hard state during the valleys of the reflares. The observations at fractional rms amplitudes $<$ 10 per cent, with count rates $>$ 100 cts s$^{-1}$, are those whose fit was dominated by the disc component, implying that J1535 transitioned to softer states (likely soft intermediate state at the peak of reflare 2 and soft state at the peak of reflare 1, where the corona was undetectable). 

\section{Discussion}
We present a detailed analysis of the spectral and timing properties of the reflaring episodes observed at the end of the discovery outburst of the black hole candidate J1535. The almost daily sampling and the low-energy coverage (down to 0.5 keV) offered by \textit{NICER}, provided an unprecedented opportunity to study these reflaring episodes in great detail. We found that during these reflares, J1535 underwent state transitions, reaching softer states at the peaks (where the Comptonised component is not significantly detected) and returning to the hard state during the valleys (where the disc component is not significantly detected). These state transitions track a hysteresis loop in the HID, resembling to some extent the hysteresis displayed in the full outburst of this and other LMXBs with both black hole \citep[e.g.][]{2010MNRAS.403...61D} and neutron star \citep[e.g.][]{2014MNRAS.443.3270M} accretors. These results support the suggestions that outbursts and reflares are driven by the same physical processes.

\subsection{Outburst reflares of MAXI~J1535-571: the state transition identification}
The HID is commonly used to trace the state of a LMXB during an outburst. \textit{NICER} started observing J1535 when the source was already bright, and did not sample the rise of the hard branch in the HID. However, a ``q-pattern" is still visible showing that J1535 underwent the hard-to-soft and the soft-to-hard transitions during the main, bright outburst \citep[see also][]{2018MNRAS.480.4443T}. The reflares in J1535 also displayed hysteresis in the HID, suggesting similar state transitions that we then demonstrated with our spectral fits. \textit{Swift/XRT} observations of these flares also suggested state transitions \citep{2019ApJ...878L..28P}; however the \textit{Swift/XRT} data quality did not allow the authors to detect the hysteresis loops.

The reflares hysteresis loops occur at 100 times lower luminosities than the main peak of the outburst. The peak luminosities of the reflares imply that the state transitions took place at L$_X\leq$ 7$\times$10$^{36}$ erg s$^{-1}$; these values are, to our knowledge, the lowest luminosity hard-to-soft transitions ever observed in a black hole binary. This also implies that the soft-to-hard transitions occurred at lower luminosity, however such low luminosities have been observed before \citep[e.g.][]{2014ApJ...791...70T,2016MNRAS.458.1636S}. We note that our luminosity estimations assumed the most likely distance value of 4.1 kpc from \citet{2019MNRAS.488L.129C}. However, these authors also estimated an upper limit for the distance of $\sim$7~kpc, which results in peak luminosities of $\leq$~2$\times$10$^{37}$~erg~s$^{-1}$ for the reflares [\citet{2019MNRAS.487.4221S} estimated a distance of 5.4$^{+1.8}_{-1.1}$ kpc, which lies between the above-mentioned values]. Although such luminosities would be among the highest observed for reflares in LMXBs, they imply state transitions occurring at more usual intensity levels.

The spectral fits yielded 2.3 $<\Gamma<$ 2.7 and 0.3~$<$~kT$_{in}<$~0.55~keV at the peak of the first 3 reflares. These values, in agreement with the soft state revealed during the main outburst \citep{2018PASJ...70...95N,2018MNRAS.480.4443T,2018ApJ...868...71S}, are consistent with the presence of an accretion disc. At the bottom of the reflares, the disc component was not significantly detected, while the fitted model yielded 1.5 $<\Gamma<$ 2. This is again in agreement with the hard state values reported at the main outburst \citep{2018PASJ...70...95N,2018MNRAS.480.4443T,2018ApJ...868...71S}. 

The state transitions observed during the reflares evidence the (partial) non-detection of the Componised component during the peaks. Since our data set and current models do not allow to determine the temperature or size of the corona during the peaks, we can only speculate that the non-detection could be due to either a drop of the temperature of the corona \citep[e.g.][]{2020ApJ...889L..18Y} or a decrease of the corona size. In the first case, the electrons would not have enough energy to upscatter the photons from the disc. Without a heating mechanism, the corona would cool down very quickly \citep{2001MNRAS.321..549M}; while it is unclear what the heating mechanism is, it is clear that it is very efficient in black hole candidates in the low-hard state, given that the temperature of the corona is quite high in those cases \citep[e.g.][]{1998MNRAS.301..435Z,2017ApJ...851..103X}. If the corona is cooler at the peaks than at the minima of the reflares, the efficiency of the heating mechanism must change rather quickly. In the second case, the reduction of the size of the corona implies a drop in the flux of the Comptonised component \citep{2019Natur.565..198K,2020MNRAS.492.1399K,2020MNRAS.494.1375Z}. In both cases, if it is hot enough, the disc could dominate the spectrum as we see at the peak of the reflares of J1535. Similarly, we speculate that the disappearance of the disc during the valleys could be caused by a cooling of the disc and an increase of its inner radius \citep[see, however,][]{2020ApJ...890...53S}. However we do not have strong evidence that this is the case.

The analysis of the X-ray variability for J1535 supports our spectral state identification  during the reflares. We found that the observations that sample the valleys (i.e. where the Comptonisation component dominates), lie along the ``hard line" of the RMS-intensity diagram plotted in Figure \ref{fig:rvi}. As J1535 becomes brighter the reflares become softer (i.e. where the disc component dominates) and the X-ray variability decreases, reaching  fractional rms amplitudes $<$ 5 per cent. It is worth noting that quasi-periodic oscillations (QPOs) were detected during the main outburst of J1535 \citep{2018AstL...44..378M,2018ApJ...865L..15S,2018ApJ...866..122H,2018ApJ...868...71S,2019ApJ...875....4S,2019MNRAS.487..928S}, however, we did not detect any significant QPO during the reflares. This result could potentially be related to differences in the  physical conditions during the normal bright outburst and the fainter reflares or it could be that QPOs are too weak to be detected at the low count rates we observe. However, the relationship between outbursts, reflares, state transitions and QPOs is probably complex. For example, Zhang et al. (submitted) reported on the spectral and X-ray variability study of the recent outburst of the black hole LMXB MAXI~J1348-630. These authors found that the source underwent a bright outburst during which, similarly to J1535, it displayed Type-B and Type-C QPOs \citep{2005ApJ...629..403C,2011MNRAS.418.2292M}. After the main outburst, MAXI~J1348-630 also underwent 3 reflares \citep{2019ATel12829....1R,2019ATel12838....1N,2019ATel13188....1Y,2020ATel13451....1P,2020ATel13459....1S,2020ATel13465....1Z}. However, contrastingly to J1535, Zhang et al. found that none of the reflares experienced state transitions (the source remained in the hard state during all flares) and they detected Type-C QPOs also during these reflares.

Finally, we found that each reflare is fainter than the preceding one (see also \citealt[][]{2019ApJ...878L..28P} and Section \ref{origin} for further discussion), albeit their duration appears to remain approximately constant ($\sim$30 days). Additionally, we observe that the disc component follows the same behaviour, as it can be inferred from both spectral and timing parameters. The evolution of the disc flux and temperature [panels (c) and (e) in Figure \ref{fig:spec_param}] mimic the shape of the light curve, while the fractional rms evolves from $<$ 5 per cent at the peak of reflare 1 to $\sim$20 per cent in reflare 4 (see Figure \ref{fig:rvi}), indicating that the disc component becomes weaker as the reflares get fainter.

\subsection{Spectral and timing analysis of reflares in other LMXBs}
In the previous Section we briefly discussed the results of Zhang et al. (submitted) on the reflares observed in the black hole candidate MAXI~J1348-630. In addition to this source, outburst reflares have been reported for other sources, both black hole \citep[e.g.][]{1996ApJ...462L..87K,1998NewAR..42....1K,2002MNRAS.334..999Z,2003ApJ...592.1100T,2013ApJ...775....9H,2017MNRAS.470.4298Y} and neutron star  \citep[e.g.][]{2010A&A...513A..71S,2016ApJ...817..100P,2019MNRAS.483.4628V,2019ApJ...877...70B} X-ray binaries. \citet{2017MNRAS.470.4298Y} studied the X-ray spectra of the black hole candidate GRS~1739-278 during the 2 reflares observed after its 2014 outburst. These reflares underwent spectral state transitions and showed hysteresis loops in the HID \citep[see also][]{2019AstL...45..127B}. \citeauthor{2017MNRAS.470.4298Y} defined the hard state as the period when a simple power law (with $\Gamma<$ 2.1) fitted the spectra, the soft state when the disc component was $>$ 80 per cent of the total flux (0.5$-$10 keV), and the intermediate state between those two. Under those definitions, these authors found that the hard-to-soft transition luminosity and the soft state peak luminosity, during both the reflares and the main outburst, follow a correlation between those luminosities that was previously observed for bright X-ray binaries \citep{2004ApJ...611L.121Y,2011RAA....11..434T}. This result led \citet{2017MNRAS.470.4298Y} to conclude that the same mechanism is driving both outbursts and reflares. 

Similar reflares to the ones observed in GRS~1739-278 were detected for the black hole systems MAXI~J1659-152 \citep{2013ApJ...775....9H} and XTE~J1650-500 \citep{2004ApJ...601..439T}. The energy spectra of both sources during the reflares were fitted by a simple power law with $\Gamma\sim$ 1.55$-$2.5 and $\Gamma\sim$~1.66$-$1.93, respectively, implying that the sources remained in the hard state and no state transition took place. Moreover, \citet{2017MNRAS.470.4298Y} re-analysed MAXI~J1659-152 and XTE~J1650-500 data and determined there was no hysteresis during their reflares. 

A reflare was also observed after the 2018 outburst of the accreting millisecond pulsar IGR~J17379-3747 \citep{2019ApJ...877...70B}. The spectral modelling  revealed that both a single power law model and a power law plus a blackbody model produced analogous results. In any case, $\Gamma\sim$ 1.9 implying the source became relatively soft, but that it did not underwent a state transition.

While the spectral analysis of reflares has been difficult, X-ray variability studies have been even more challenging. These types of analyses were mostly reported in a few cases for individual observations taken during a reflare \citep[e.g.][]{2002PASJ...54..609T,2004ApJ...601..439T,2005MNRAS.363..112M,2017MNRAS.468.4748C}. In addition, a more extensive study of the X-ray variability during reflares was performed for the accreting millisecond pulsar SAX~J1808.4-3658. This source exhibited reflares after 5 outbursts and the presence of a 1~Hz QPO was observed after 3 of them \citep{2000IAUC.7358....3V,2001ApJ...560..892W,2009ApJ...707.1296P}, which was interpreted as being caused by hydrodynamic disc instabilities. However, the later report of a similar type of variability in the same source, but at higher luminosities, suggests that these QPOs are not related to the flaring state \citep{2014ApJ...789...99B}. Likewise, \citet[][]{2019ApJ...877...70B} analysed the X-ray variability in the accreting millisecond pulsar IGR~J17379-3747. In particular, these authors were able to study the properties of the coherent pulsations during both the main 2018 outburst and the subsequent reflares. Table \ref{summary} summarises the main results from the spectral and timing studies during the reflares of the sources discussed above.

\begin{table*}
\caption{Summary of LMXBs with spectral and/or timing studies during reflares}
\begin{tabular}{ccccc}
\hline 
Source & State transitions & Hysteresis & QPOs & Reference \\
\hline
MAXI~J1535-571 & yes & yes & no & This work \\
MAXI~J1348-630 & no & no & yes & Zhang et al. (submitted) \\
GRS~1739-278 & yes & yes &  & \citet{2017MNRAS.470.4298Y,2019AstL...45..127B} \\
MAXI~J1659-152 & no & no &  & \citet{2013ApJ...775....9H} \\
XTE~J1650-500 & no & no &  & \citet{2004ApJ...601..439T} \\
IGR~J17379-3747 & no &  & no & \citet{2019ApJ...877...70B} \\
SAX~J1808.4-3658 &  &  & yes & \citet{2000IAUC.7358....3V,2001ApJ...560..892W}; \\
 &  &  &  & \citet{2009ApJ...707.1296P} \\
\noalign{\vskip 1mm}
\hline
\end{tabular}
\label{summary}
\end{table*}

Thanks to the low-background high-sensitivity of \textit{NICER} and its capabilities of performing monitoring observations, it is now possible to do X-ray variability studies of low-luminosity outburst reflares like the one presented here and in Zhang et al. (submitted) for MAXI~J1348-630. With the current data (two outburst of two different sources) it is not possible to discern whether there is a link between QPO occurrence and the spectral evolution of the reflares. This highlights the need for more monitoring observations of these reflares with instruments like \textit{NICER}.

\subsection{Origin of reflares}
\label{origin}
The disc instability model (DIM) is the model most commonly used to explain how outbursts start \citep[for a review see][]{2001NewAR..45..449L}. In this model, the outburst is the result of the heating of the disc due to the ionisation of the hydrogen accumulated during the quiescent state. The DIM is not able to consistently explain the onset of reflares, as the model implies that the disc is depleted from matter at the end of an outburst, while the reflares require a large amount of matter left \citep[see][and references therein]{2016ApJ...817..100P}. Although it is not clear whether the different types of reflares are produced by the same physical process, their origin was discussed several times in the literature. 

\begin{itemize}
    \item \citet{1993ApJ...408L...5C} proposed a scenario that generates two glitches or reflares (according to the classification in \citealp{2019ApJ...876....5Z}) after the outburst, in which irradiation of the secondary star leads to a mass transfer process. The first maximum would be caused by X-ray evaporation of the material close to the first Lagrangian point, while the second one would be the result of mass transfer instability due to the heating of the secondary during the main outburst. 
    \item Other authors have favoured the enhanced mass transfer from the secondary caused by X-ray heating. For instance, \citet{1993A&A...279L..13A} showed that if this is the case, then the main outbursts are caused by a burst of mass transfer from the secondary. 
    \item \citet{1998MNRAS.301..382S} proposed that single reflares are produced when irradiation causes an increase in viscosity in the outer part of the disc, generating a mass front that propagates inward \citep[see also][]{1998MNRAS.293L..42K,1998NewAR..42....1K}.
\end{itemize}

Reflares are very common in dwarf novae systems. Particularly in the SU UMa systems, the reflares are only observed after long outbursts, known as superoutbursts. 
It was proposed that the reflares could be caused by enhanced mass transfer as a result of illumination of the donor \citep{2000NewAR..44...15H,2000A&A...353..244H}, however this enhancement was proven to be negligible \citep{2003A&A...401..325O,2004A&A...428L..17O}. 
\citet{2015PASJ...67...52M} proposed that the formation of magnetic fields during the superoutburst results in an enhanced viscosity, which produces wave reflections between the hot and cold states of the accretion disc, causing the reflares. \citet{2019ApJ...876....5Z} proposed that mini-outbursts in black hole systems could be explained by the DIM if the disc is still hot at the end of the outburst, as is the case of dwarf novae systems. Usually single or multiple reflares are observed in dwarf novae systems and, in the later case, they display a periodic behaviour. Although is not the usual behaviour for LMXBs, the nearly periodic occurrence of the reflares of J1535 resemble those observed in dwarf novae. Moreover, the reflaring light curves we obtain for J1535 are similar to the one observed at optical wavelengths in GRO~J0422$+$32 \citep{1996ApJ...462L..87K} and MAXI~J1807$+$132 \citep{2019MNRAS.484.2078J}.
\citet{2019ApJ...876....5Z} noticed that the extrapolation of the decaying phase of the main outburst in the light curve coincides with the peaks of the reflares for both dwarf novae and LMXBs, implying the existence of a critical temperature of the disc.

While plausible, all the above models proposed to explain reflares in LMXBs are unproven. If we consider, for example, that reflares are triggered by the matter left in the disc after the main outburst, it remains unknown whether each reflare is powerful enough as to trigger the next one. Additionally, one can wonder whether mass transfer from the secondary can trigger the reflares in time scales of $\sim$30 days, or if reflares can result from wave reflections through the disc so regularly.

Independently of the trigger, \citet{2016ApJ...817..100P} proposed that the reflares are small-scale outbursts, generated by a disc instability (a change in the disc density) at the end of an outburst. Although what triggers the new instability is still unclear, the lower scale appears to be due to the low amount of matter left in the disc after the main outburst. Likewise, the progressive faintness of the reflares in J1535 supports the interpretation of an emptying reservoir of mass available for accretion, as proposed by \citet{2019ApJ...878L..28P}. Additionally, these authors detected a compact radio jet in the hard state of the reflares that was observed to quench as the source transitioned to the soft state. This behaviour provides further support to the conclusion that the relevant physical ingredients of the main outburst are very similar to those describing the reflares.

In summary, as it was pointed before by several authors, reflares are observed in different accreting systems (dwarf novae, neutron stars, black holes), which indicates that they are related to the accretion process and independent of the nature of the compact object. Moreover, the reflares observed in J1535 showed state transitions and hysteresis, as is usually observed during bright outbursts of LMXBs. This result suggests that similar physical processes take place during outbursts and reflares, even when the X-ray luminosity is two orders of magnitude different.
%

\section{Conclusions}

In this work we present a detailed analysis of the evolution of the three brightest outburst reflares observed in the black hole candidate MAXI~J1535-571 during the decline of its discovery outburst. 
Both our spectral and X-ray variability results show that during the reflares the source underwent state transitions, tracking hysteresis loops. 
J1535 reached the soft state at the peak of the first reflare, while it returned to the hard state during the valleys. We found that these state transitions were characterised by the appearance and disappearance of the disc component, implying that both the disc and the corona change significantly. 
Assuming the most probable distance value of $\sim$4 kpc, we found that the state-transition X-ray luminosities are the lowest luminosities ever observed at which the hard-to-soft state transitions took place and among the lowest at which the soft-to-hard state transitions ever happened.
The similarities observed between the reflares and the main outburst of this and other black hole binaries support the interpretation that reflares are smaller scale outbursts, driven by the same physical mechanisms.

\section*{Acknowledgements}
V.A.C. acknowledges support from the Royal Society International Exchanges ``The first step for High-Energy Astrophysics relations between Argentina and UK" and from the Spanish \textit{Ministerio de Ciencia e Innovaci\'on} under grant AYA2017-83216-P. 
K.A acknowledges support from a UGC-UKIERI Phase 3 Thematic Partnership
(UGC-UKIERI-2017-18-006; PI: P. Gandhi).
L.Z. acknowledges support from the Royal Society Newton Funds.
D.A. and D.J.K.B. acknowledge support from the Royal Society. 
J.A.C was supported by PIP 0102 (CONICET) and PICT-2017-2865 (ANPCyT). This work was also supported by the \textit{Agencia Estatal de Investigaci\'on} grant AYA2016-76012-C3-3-P from the Spanish \textit{Ministerio de Ciencia e Innovaci\'on} and by the \textit{Consejer\'ia de Econom\'ia, Innovaci\'on, Ciencia y Empleo of Junta de Andaluc\'ia} under research group FQM-322, as well as FEDER funds.
T.M.D. acknowledges support via the Ram\'on y Cajal Fellowship RYC-2015-18148.
This work was supported by NASA through the \textit{NICER} mission and the Astrophysics Explorers Program. This work made use of data supplied by the UK Swift Science Data Centre at the University of Leicester. This research has also made use of data and software provided by the High Energy Astrophysics Science Archive Research Center (\textsc{heasarc}) and NASA's Astrophysics Data System Bibliographic Services.




\bibliographystyle{mnras}
\bibliography{maxij1535} 







\bsp	
\label{lastpage}
\end{document}